\documentstyle[preprint,eqsecnum,aps]{revtex}
\begin{document}
\preprint{}
\title{Electron and Quasiparticle Exponents of Haldane-Rezayi state 
in Non-abelian
Fractional Quantum Hall Theory }
\author{Jen-Chi Lee\footnote{ 
On leave of absence from : Department of Electrophysics, National
Chiao-Tung University, HsinChu, Taiwan, ROC.}
and Xiao-Gang Wen}
\address{Department of Physics, MIT, 77 Massachusetts Avenue, Cambridge, MA
02139, USA}
\maketitle
\begin{abstract}
The quasiparticle propagator of Haldane-Rezayi(HR) fractional quantum
Hall (FQH) state is calculated, based on a chiral fermion model
(or a Weyl fermion model) equipped with a 
hidden spin $SU(2)$ symmetry. The spectrum of the chiral fermion model
for each {\em total spin} and total momentum is shown to be identical to
that of the $SU(2)$ $c=-2$ model introduced to describe the edge
spectrum of the HR state.
\end{abstract}
\pacs{}

\section{Introduction}

Incompressible quantum Hall liquids form a new class of matter states which
contain non-trivial topological orders. It is known that quantum Hall
states can be divided into two classes: abelian quantum Hall state
and non-abelian quantum Hall state.\cite{revW} Almost all observed
quantum Hall states are abelian states. Although the
observed filling fraction $\nu=5/2$ state may be a non-abelian
states, we still have no direct evidence of non-abelian state in experiments.
To confirm a non-abelian state in experiments, we need to use
tunneling experiments to measure
exponents of electron (or quasiparticle) propagator along the edge.

Many non-abelian incompressible FQH fluids have been the subject of active
theoretical
research for the last few years. The notion of non-abelian quantum Hall states
was first proposed by Moore and Read \cite{a}, 
largely motivated by an observation that
many-body FQH wave functions can be constructed as a correlation function
in a 2d conformal field theory (CFT).  Later non-abelian quantum Hall 
states were also constructed through parton construction
that leads to effective theories with non-abelian
gauge field.\cite{b} However the two approaches
were shown to be closely related to each other.\cite{c}
The non-abelian states are characterized by the existence of non-abelian 
quasiparticles. Although theoretically we still cannot
prove the non-abelian Berry's phases induced by exchanging two quasiparticles,
some evidences for their non-abelian statistics have recently been tested and
confirmed through the counting of number of degeneracy of quasiparticle 
excitations.$\cite{d}$

The non-abelian states can also be characterized (more completely)
through their edge structure. This characterization allows us to
identify non-abelian states in experiments.
The edge excitations of any abelian states
can be described by edge phonon theory with several branches 
(more precisely, by
several Gaussian models or by several $U(1)$ current algebras).
While the edge excitations of a non-abelian state is described by a theory
which cannot be identified as the Gaussian models. The abelian and non-abelian 
states have different sets of 
exponents for electron and quasiparticle propagators.

The edge theory for a non-abelian state can be obtained through the 
parton construction that produces the non-abelian bulk state.\cite{b}
It can also
be obtained through a (conjectured) relation between the bulk CFT that 
produces the bulk wave function and the edge CFT that describes the edge
excitations.\cite{e} For all known abelian and non-abelian states,
the edge CFT was always identical to the minimal bulk CFT.\cite{f,g}
It was shown in Ref. \cite{f} (under certain assumptions) that
the edge CFT (which is identified with the minimal bulk CFT) always
reproduces the low energy spectrum of edge excitations.
In many case (such as abelian states and the Pfaffian state),
the natural inner product of the  edge CFT is also identical to
the inner product of edge excitations defined through the electron
wave functions. (This is one of big unsolved mysteries, and there are
exceptions.)
For those quantum Hall states, we may use the known correlations in the
edge CFT to calculate electron and quasiparticle propagators along the edge.
Both edge spectra and the electron propagators obtained from edge CFT
have been confirmed by direct numerical calculations for several
non-abelian states.

One of the most important non-abelian states is the 
HR state for spin 1/2 electrons
$\cite{h}$. The HR state was proposed to explain the plateau at filling
fraction $\nu = 5/2$ observed in experiment $\cite{i}$  and thus may serve as
the first candidate of real non-abelian FQH states. There are two approaches for
the HR state. One is based on the 
$SU(2)$ $c = -2$ CFT ( plus an abelian U(1) Gaussian
part).\cite{f,g} Comparing to the minimal $c=-2$ CFT,
the $SU(2)$ $c=-2$ CFT has an additional $SU(2)$ symmetry and primary fields
of dimension $h_s=\frac18[(4s+1)^2-1]$
which form an irreducible representation of $SU(2)$ with spin $s$.
In the rest of the paper, we will simply refer the $SU(2)$ $c = -2$ CFT
as the $c = -2$ CFT.
The edge spectrum can then
be analyzed by the known character of CFT in the $c = -2$ model. However, due to
its nonunitarity, the $c = -2$ model has negative-norm states in its
Hilbert space if one uses its natural inner product. 
Presumably, the inner product between physical edge states and that 
between the CFT states are not the same. 
Although this drawback does not
ruin the CFT description of the spectrum, it does prohibit us from doing any
further calculation, 
such as electron and quasiparticle propagators along the edge,
based on the CFT techniques. 
The other
approach  \cite{j} uses a free field realization of the $c=-2$ CFT,
with a lagrangian $L=|\partial_t \eta|^2 - |\partial_x \eta|^2$ where
$\eta$ is a complex fermionic field. The standard quantization leads
to the $c=-2$ theory with negative-norm states. However its was shown in 
Ref.~\cite{j} that one can redefine the inner product which
remove all negative-norm states and {\em leave the correlations
between electron operators unchanged}. Thus the new theory with
the redefined inner product can still reproduces the bulk wave function
of HR state through correlations of electron operators.
In this way, the exponent of electron propagator is found to be 4
which agrees with value obtained form CFT.\cite{g}.
The exponent of quasiparticle propagator is still unknown. If we had used
CFT results, the exponent would be $-1/8$, which is physically
unacceptable.  In this paper, we will calculate the exponent of 
quasiparticle propagator and the exponent will be
found to be $+3/8$. We will also
present some numerical results that support the exponent 4 for the
electron propagator.

The Vector space of the $c=-2$  CFT is identical to that of a chiral
fermion theory (or, in string theory terminology, a 
Weyl fermion or two Majorana-Weyl fermions).\cite{j,mapL}
After the redefinition
of the inner product for the $c=-2$ CFT, the two theories even have the
same inner product.\cite{j,mapL} The authors of Ref.~\cite{cfC}
used the chiral fermion theory to describe the edge excitation of HR state
and found the electron exponent to be 3. This result disagrees with
the result from Ref.~\cite{j,g}, and is not favored by our numerical
result.

In section two we show that the low energy edge excitations of the HR state
generated by the $c = -2$ CFT is equivalent to those generated 
by a chiral fermion (in the twisted sector). 
In section three we identify the electron operators 
constructed in $\cite{j}$  and calculate their
correlation function, which
reproduces the HR wave function. 
We then use these electron operators to define quasiparticle operators
by imposing the appropriate boundary conditions. This allows us
calculate the exponent of quasiparticle propagator.
No CFT calculation was used. 
Finally we generalized
this technique to other non-abelian FQH states.
The quasiparticle exponents calculated based on this technique
are consistent with known results for those non-abelian FQH states.

\section{the edge spectrum}

The HR state is a d-wave paired spin singlet FQH state
with filling fraction $\nu=1/m$:
\begin{eqnarray}
\Phi_{HR} (z_i,w_i)& = & \Phi_{m}\Phi_{ds}(z_i,w_i), \nonumber\\
\Phi_{ds}&  = &{\cal A}_{z,w}(\frac{1}{(z_1-w_1)^2}
\frac{1}{(z_2-w_2)^2}...),\nonumber\\
\Phi_{m} & = &( \prod_{i<j} (z_i-z_j)^m \prod_{i<j} (w_i-w_j)^m \prod_{i<j}
(z_i-w_j)^m ) \exp(-\frac14 \sum_{i}(|z_i|^2 + |w_i|^2) ),
\label{0}
\end{eqnarray}
where $ {\cal A}_{z,w}$ is the anti-symmetrization operator which performs 
separate anti-symmetrizations among  $z_i$'s and among $w_i$'s ;  
$m$ is an even integer. Here $
z_i (w_i)$  are the coordinates of the spin-up (-down) electrons. 
The edge spectrum of a circular droplet of HR state 
is labeled by total angular momentum 
and forms a representation of the spin $SU(2)$.
Let $N^{nab}_{L,s}$ be 
the number of edge excitations 
for the non-abelian part (we will ignore the abelian U(1) part)
which carry the following quantum numbers: the total 
angular momentum is $L$, the total spin is $s$ and the $z$-component
of the total spin is $S_z=0$ (or $1/2$ if $s$ is a half-odd-integer).
$N^{nab}_{L,s}$ was described in 
Ref.~\cite{g} by the following character in the $c = -2$ model
\begin{equation}
Ch_s(\xi) \equiv \sum_{n=0}^\infty N^{nab}_{L_0+n,s} \xi^{n}
= \frac{\xi^{h_s} - \xi^{h_{s+1/2}}}{\prod_{n=1}^\infty (1 - \xi^n)}
\label{1}
\end{equation}
where $ h_s = \frac18[(4s+1)^2-1]$ and $L_0$ is the total angular momentum
of the ground state in the spin $s=0$ sector. 
Multiplying Eq.  (\ref{1}) by $(2s+1)$
gives us number of edge states in each total spin $s$ sector. On the other hand,
we can also use the following Hamiltonian proposed in Ref.~$\cite{j}$
\begin{eqnarray}
H &=& v\sum_{k=1}^{\infty} k (c^\dagger_{k\uparrow}c_{k\uparrow}
+c^\dagger_{k\downarrow}c_{k\downarrow}),\nonumber\\
\{c^\dagger_{k\sigma},c_{k^\prime\sigma^\prime}\} & = &
\delta_{kk^\prime}\delta_{\sigma\sigma^\prime}, \nonumber\\
 \{c_{k\sigma},c_{k^\prime\sigma^\prime}\} &=&
\{c^\dagger_{k\sigma},c^\dagger_{k^\prime\sigma^\prime}\} = 0 
\label{2}
\end{eqnarray}
to describe the edge excitations of HR state.
Here $v$ is the edge velocity ( which will be set to $v = 1$ in the following
discussion) and we have put the system on a circle $ x \in [0,2\pi)$. 
The Hamiltonian in Eq. $(\ref{2})$ has a global $SU(2)$
symmetry with the total spin generators $\cite{j}$
\begin{eqnarray}
S^z & = & \frac12 
\sum_{k=1}^{\infty} (c^\dagger_{k\uparrow}c_{k\uparrow} - c^\dagger
_{k\downarrow}c_{k\downarrow}),\nonumber\\
S^+ & = & 
\sum_{k=1}^{\infty} c^\dagger _{k\uparrow}c_{k\downarrow},\nonumber\\
S^- &  = & 
\sum_{k=1}^{\infty} c^\dagger _{k\downarrow}c_{k\uparrow}.
\label{7}
\end{eqnarray}

We have tabulated the spectrum of Eq. $(\ref{2})$ for up to 6th low-lying
energy eigenstates for each spin sector with $ s \leq 3/2$. All results are the
same as those generated by the character in Eq. $(\ref{1})$. 
To prove a general relation,
let $N_{K,s}$ be number of states of Eq. $(\ref{2})$ with total
momentum $K$, total spin $s$ and total $z$-component of spin $S_z=0$. 
Let $n_{K,M}$ be number of states
with total momentum $K$ and total $z$-component of spin $S_z=M$.
It is clear that  $N_{K,s}=n_{K,s}-n_{K,s+1}$.
The character for
$n_{K,M}$ can be calculated:
\begin{eqnarray}
ch(\eta,\xi)&\equiv& \sum_{K,M} n_{K,M} \eta^{2M} \xi^K  \nonumber \\
&=& \prod_{k=1}^\infty (1+\eta \xi^k) (1+\eta^{-1} \xi^k)
\label{ch}
\end{eqnarray}
To evaluate the product in Eq.~(\ref{ch}), we consider the follow 
(twisted) chiral fermion model
\begin{equation}
H=\sum_{k=0,\pm 1,\pm 2,...} k :c^\dagger_k c_k: =
\sum_{k=1,2,...} k c^\dagger_k c_k + 
  \sum_{k=-1,-2,...}  (- k) c_k c_k^\dagger
\label{cf}
\end{equation}
We note that the model described by Eq.~(\ref{cf}) is identical to that
described by Eq.~(\ref{2}) except the $k=0$ mode.
Let $n^c_{K,Q}$ be number of states
with total momentum $K$ and total fermion number $Q$, where $Q$
is given by
\[
Q=\sum_{k=1}^\infty c^\dagger_k c_k  
- \sum_{k=0}^{-\infty} c_k c^\dagger_k
\]
Note that an empty $k=0$ level is regarded as the presence of a hole
which contribute $-1$ to the total fermion number $Q$.
Through bosonization
we know that $n^c_{K,Q}$ is equal to the partition number $p_{K-K_0}$
where $K_0(Q)=(Q+1/2)^2/2-1/8$ 
is the minimum momentum of states with $Q$ fermions.
Using $\sum_{i=0}^\infty p_i \xi^i=1/\prod_{n=1}^\infty (1-\xi^n)$, we find
\begin{eqnarray}
ch^c(\eta,\xi)&\equiv& \sum_{K,Q} n^c_{K,Q} \eta^Q \xi^K  \nonumber \\
&=& (1+\eta^{-1})\prod_{k=1}^\infty (1+\eta \xi^{k}) (1+\eta^{-1} \xi^{k})\\
&=& \sum_Q \frac{\xi^{K_0(Q)} \eta^Q}{\prod_{n=1}^\infty (1-\xi^n) }
\label{chc}
\end{eqnarray} 
The $(1+\eta^{-1})$ term is the character for the $k=0$ level.
Compare Eq.~(\ref{ch}) with Eq.~(\ref{chc}), we see that
\begin{eqnarray}
ch(\eta,\xi)&=&  ch^c(\eta,\xi)/(1+\eta^{-1}) \nonumber \\
&=& \sum_Q \frac{\eta^Q \sum_{n=0}^\infty (-)^n \xi^{K_0(Q+n)} 
}{\prod_{n=1}^\infty (1-\xi^n) }
\end{eqnarray}
After replacing $Q$ by $2M$ and noticing that
$K_0(2M)=h_{M}$, we find
\begin{eqnarray}
ch_M(\xi)&\equiv& \sum_{K} n_{K,M} \xi^K  \nonumber \\
&=& \frac{ \sum_{n=0}^\infty (-)^n \xi^{h_{M+\frac{n}{2} }}
}{\prod_{n=1}^\infty (1-\xi^n) } 
\end{eqnarray}
From the relation $N_{K,s}=n_{K,s}-n_{K,s+1}$ we find
\begin{eqnarray}
Ch_s(\xi)&=&ch_s(\xi)-ch_{s+1}(\xi) = \frac{ \xi^{h_s} - \xi^{h_{s+1/2 } }
}{\prod_{n=1}^\infty (1-\xi^n) }
\end{eqnarray} 
which is exactly the character obtained from the $c=-2$ CFT.
Thus the spectrum of Eq.~(\ref{2}) labeled by (angular) momentum and the total
spin is identical to that of $c=-2$ CFT labeled by the same quantum number.

\section{the electron and quasiparticle exponents}

Finding the theory with positive definite inner product
that completely describes the spectrum of edge
excitations is not the end of story.
We also need to identify the electron
operators in the theory. One naive guess would be $
\psi_{\sigma}e^{i\sqrt{m}\phi}$, where
\begin{eqnarray}
\psi_{\uparrow}(t,x)& = &\sum_{k=1}^{\infty } c_{k,\uparrow}
e^{-ik(t-x)} + \sum_{k=1}^{\infty} c^\dagger _{k,\downarrow}
e^{ik(t-x)},\nonumber\\
\psi_{\downarrow}(t,x)& = &\sum_{k=1}^{\infty}
c_{k,\downarrow}e^{-ik(t-x)} - \sum_{k=1}^{\infty} c^\dagger
_{k,\uparrow}e^{ik(t-x)}
\label{psi0}
\end{eqnarray}
This is the choice made in Ref.~\cite{cfC}. The correlation of 
$\psi_{\sigma}e^{i\sqrt{m}\phi}$ contains two parts. The
first part comes from $e^{i\sqrt{m}\phi}$ which reproduces the
abelian part $\prod (z_i-z_j)^m(z_i-w_j)^m(w_i-w_j)^m$
of the HR wave function. But the second part coming from $\psi_{\sigma}$
fails to reproduce the non-abelian part 
${\cal A}_{z,w}( \frac{1}{(z_1 - w_1 )^2} \frac{1}{(z_2 - w_2 )^2}...)$
of the HR wave function. In this section
we will follow the following
principle to choose the electron operators. We will require
the electron operators to reproduce the bulk wave function.
Such a principle, although has not been proven to be correct,
has led to correct electron operators for abelian states and 
for a non-abelian Pfaffian state. This principle leads us to the
electron operator introduced in Ref.~\cite{j}: 
\begin{equation}
\psi_{e,\sigma}e^{i\sqrt{m}\phi}
\label{10}
\end{equation}
where
\begin{eqnarray}
\psi_{e \uparrow}(t,x)
& = &\sum_{k=1}^{\infty } \sqrt{k}c_{k,\uparrow}
e^{-ik(t-x)} + \sum_{k=1}^{\infty} \sqrt{k}c^\dagger _{k,\downarrow}
e^{ik(t-x)},\nonumber\\
\psi_{e \downarrow}(t,x)
& = &\sum_{k=1}^{\infty}
\sqrt{k} c_{k,\downarrow}      e^{-ik(t-x)} - \sum_{k=1}^{\infty} 
\sqrt{k} c^\dagger_{k,\uparrow}e^{ ik(t-x)}
\label{9}
\end{eqnarray}
The correlator of $\psi_{e,\sigma}$ can then be calculated to be 
\begin{equation}
< \psi_{e\uparrow}(z_1)
\psi_{e\downarrow}(w_1) \psi_{e\uparrow}(z_1)
\psi_{e\downarrow}(w_1)...>
 =  {\cal A}_{z,w}( \frac{1}{(z_1 - w_1 )^2} \frac{1}{(z_2 - w_2 )^2}...),
\label{11}
\end{equation}
which reproduces the non-abelian part of the HR wave function. 
We will take Eq. $(\ref{10})$   as
the electron operators near the edge. The electron propagator on the edge thus
has the following form
\begin{eqnarray}
G_{e} (t,x) \sim \frac{1}{(x-vt)^{g_e}},\ \ \ \ \    g_{e} = m + 2 .
\label{12}
\end{eqnarray}
When $m=2$, such a propagator with exponent $g_e=4$
leads to a set of occupation numbers that satisfies
\begin{equation}
n_{l_0}:n_{l_0-1}:...=1:4:10:20:35:...
\end{equation}
Here $n_l$ is the occupation number
on the orbit with angular momentum $l$ of a circular droplet, and
$l_0$ is the last orbit occupied by electrons ({\it ie} $n_l=0$ for $l>l_0$).
Direct numerical calculation for three spin down and three spin up
electrons gives
$1:4.351:10.71:15.99:... $ and for four spin down and four spin up
electrons gives      
$1:4.299:10.39:18.52:24.68:...  $. 
If $g_e=3$ we would have $1:3:6:10:...$, and if $g_e=5$ we would have 
$1:5:15:35:...$.
We see that numerical  results almost rule out the possibility
of $g_e=3$ and $g_e=5$, and give strong support for $g_e=4$

We now turn to consider the quasiparticle propagator of HR state. 
The quasiparticle operator has a form $\eta_q e^{i\phi/2\sqrt{m}}$.\cite{g}
The non-abelian part $\eta_q$ creates a cut of $-1$ for the operator
$\psi_{e,\sigma}$, {\it ie} $\psi_{e,\sigma}$ changes sign as it
goes around $\eta_q$. Note that the abelian part $e^{i\phi/2\sqrt{m}}$
also create a cut of $-1$ to 
$e^{i\phi\sqrt{m}}$, the abelian part in the electron operator.
Thus the total quasiparticle operator $\eta_q e^{i\phi/2\sqrt{m}}$
does not generate any cut to the total electron operator
$\psi_{e,\sigma} e^{i\phi\sqrt{m}}$, as required by the single-valueness
of the electron wave function. It was also shown in Ref.~\cite{g}
that the quasiparticle carries $1/2m$ charges and zero spin. 
In the follow we will concentrate on non-abelian part of quasiparticle operator.
First we write down quasiparticle operators using the
prescribed twisted boundary condition between the 
non-abelian parts of electron and quasiparticle operators.
Introduce $ U(x_1 ,x_2)$ be the product of two $\eta_q$ fields
at equal time: $ U(x_1 ,x_2) = \eta_q(x_1) \eta_q(x_2)$.
Such an operator $ U(x_1 ,x_2)$ can be defined through its commutator
with $\psi_{e,\sigma}$
\begin{eqnarray}
U(x_1, x_2) \psi_{e,\sigma}(x) = f(x) \psi_{e,\sigma}(x)
U(x_1, x_2) 
\label{13}
\end{eqnarray}
where $f(x)$ is defined to be
\begin{eqnarray}
f(x) = \left\{ \matrix{- 1 ,& x_1<x<x_2 \cr 
                       + 1 ,& \hbox{otherwise} \cr} \right.
\label{14}
\end{eqnarray}
The quasiparticle propagator (at equal time) is then given by
\begin{eqnarray}
<0| U(x_1,x_2)| 0>.
\label{15}
\end{eqnarray}

As an example, we first work on the case of a  chiral fermion  model
\begin{eqnarray}
\psi(t,x) = \sum_{k \in Z} c_{k} e^{-ik(t-x)},
\label{15-1}
\end{eqnarray}
which corresponds to two-critical-Ising model. Since $\psi(x)$ has the
standard mode expansion, we will do it on the coordinate space. To solve
Eq. $(\ref{13})$, we first write it in the following form
\begin{eqnarray}
U_{2I}^{\alpha}(x_1, x_2)  \psi (x) 
U_{2I}^{\alpha, -1} (x_1,x_2) = e^{ i
\frac{\pi}{2} \alpha (f(x) - 1)}  \psi (x)
\label{16}
\end{eqnarray}
where $\alpha$ is a real parameter. For $\alpha = 1$ one gets Eq.
$(\ref{13})$. Since
\begin{eqnarray}
[ \psi (x) , \frac14 \int dy \psi^\dagger (y) \psi (y)(f(y) - 1) ] 
=  \frac{\pi}{ 2} (f(x) - 1) \psi(x)
\label{18}
\end{eqnarray} 
we find
\begin{eqnarray}
U_{2I}(x_1,x_2) = 
\exp\left(\frac{i}{4} \int dx\psi^\dagger (x) \psi (x)(f(x) - 1)\right).
\label{17}
\end{eqnarray}
The quasiparticle propagator can now
be easily calculated by using bosonization rule by noting
$\psi^\dagger (x) \psi (x)$ is $2\pi$ times the fermion density:
$\psi^\dagger (x) \psi (x)=2\pi\rho=\partial \phi$. We find 
$U_{2I} (x_1,x_2) =e^{i(\phi(x_1)-\phi(x_2))/2}$ and
\begin{equation}
<0|U_{2I} (x_1,x_2) | 0> = (x_1 -x_2 )^{-1/4}.
\label{19}
\end{equation}
The exponent in Eq. $(\ref{19})$ is consistent with the result of
two-critical-Ising model by using the standard CFT calculation.

We now apply Eq. $(\ref{13})$ to the case of operators $\psi_{e,\sigma}$.
We will use a different approach to calculate the average $<0|U(x_1,x_2) |0>$.
First we note that $\psi_{e \uparrow}$ can be written as
\begin{equation}
\psi_{e \uparrow}=
\sum_{k=1}^{\infty } \sqrt{k} c_{k,\uparrow}
e^{-ik(t-x)} + \sum_{k=-1}^{-\infty} \sqrt{-k} c_{k,\uparrow}
e^{-ik(t-x)} 
\end{equation}
if we rename $c^\dagger_{k,\downarrow}$ by $c_{-k,\uparrow}$ for $k>0$.
The ground state $|0>$ can be regarded as a filled Fermi sea
with negative momentum states occupied by spin $\uparrow$
electron, {\it ie}  $c^\dagger_{k,\uparrow}|0>|_{k<0}=0$.
Using the first-quantized picture, we write
\begin{equation}
<0|U(x_1,x_2)|0>=\lim_{M\to \infty}
<null| \left(\prod_{m=1}^M c_{-m,\uparrow}\right) U(x_1,x_2)
\left(\prod_{m=1}^M c^\dagger_{-m,\uparrow}\right) |null>
\label{20}
\end{equation}
where $ |null>$ is the first-quantized vacuum with no fermions, 
and $M$ is a momentum cutoff. 
According to Eq.
$(\ref{13})$, we know the operation of $ U(x_1,x_2) $ on 
$ c^\dagger_{-n ,\uparrow} = c_{n ,\downarrow} $ to be
\begin{eqnarray}
& &U (x_1,x_2) c^\dagger_{-n ,\uparrow}=U (x_1,x_2) c_{n, \downarrow} 
\nonumber \\
&= &
\sum_{m>0} \sqrt{\frac{m}{n}} f_{nm} c_{m, \downarrow} U(x_1,x_2)
-\sum_{m<0} \sqrt{\frac{m}{n}}
f_{nm} c^\dagger_{-m, \uparrow} U(x_1,x_2)  \nonumber \\
&= &
\sum_{m>0} \sqrt{\frac{m}{n}} f_{nm} c^\dagger_{-m, \uparrow} U(x_1,x_2)
-\sum_{m<0} \sqrt{\frac{m}{n}}
f_{nm} c^\dagger_{-m, \uparrow} U(x_1,x_2) 
;
\label{21}
\end{eqnarray}
whereas the operation of $ U_{2I}(x_1,x_2) $ on $ c^\dagger_{-m}$ is
\begin{eqnarray}
U_{2I} (x_1,x_2) c^\dagger_{-n} = \sum_{m} f_{nm}
c^\dagger_{-m} U_{2I}(x_1,x_2),
\label{21-1}
\end{eqnarray}
where $ f_{nm}(x_1,x_2) $ is defined to be
\begin{eqnarray}
f_{nm} = \frac{1}{2\pi} \int dx e^{i(n-m)x} f(x).
\label{22}
\end{eqnarray}
Eq.~$(\ref{20})$  can then be calculated by using Eq. $(\ref{21})$: 
\begin{equation}
<0|U(x_1,x_2)|0>= \hbox{det}( \sqrt{\frac{m}{n}} f_{nm} )
=  \hbox{det}(  f_{nm} )= <0|U_{2I}(x_1,x_2)|0>
\label{23}
\end{equation}
where $ ( \sqrt{\frac{m}{n}} f_{nm} )$ and $ (  f_{nm} )$ are $M\times M$
matrices
whose matrix elements are given by $\sqrt{ \frac{m}{n} }f_{nm}$ and $f_{nm}$,
$m,n=1,...,M$.
In Eq.~(\ref{23}) we have used the fact that $\hbox{det}(  f_{nm} )$
is just \hbox{$<0|U_{2I}(x_1,x_2)|0>$} expressed in the first-quantized picture.
We also have used 
\begin{equation}
<null| U(x_1,x_2) |null>=\hbox{const.}
\end{equation}
This is because $|null>$ is a state with no fermions
and hence $U(x_1,x_2)$ cannot affect $|null>$.
The exponent calculated this way
is thus the same as that of Eq. $(\ref{19})$.
The total quasiparticle propagator is thus
\begin{eqnarray}
G_q (t,x) \sim \frac{1}{(x-vt)^{g_q}} ,\ \ \ \ \
g_q = \frac{1}{4}+\frac{1}{4m}.
\label{24}
\end{eqnarray}
where $\frac{1}{4m}$ is the contribution from the abelian part
of the quasiparticle operator.
The exponents $g_e$ in Eq. $(\ref{12})$ and $g_q$ in Eq. $(\ref{24})$
can be measured in tunneling experiment.

\section{quasiparticle exponents of other FQH states}

The technique we have used to calculate the quasiparticle exponent can be
generalized to many other non-abelian FQH states.
First let us consider the generalized HR state given by
\begin{eqnarray}
\Phi_{HR} (z_i,w_i)& = & \Phi_{m}\Phi_{ds}(z_i,w_i), \nonumber\\
\Phi_{ds}&  = &{\cal A}_{z,w}(\frac{1}{(z_1-w_1)^{2l+2}}
\frac{1}{(z_2-w_2)^{2l+2}}...),\nonumber\\
\Phi_{m} & = &( \prod_{i<j} (z_i-z_j)^m \prod_{i<j} (w_i-w_j)^m \prod_{i<j}
(z_i-w_j)^m ) \exp(-\frac14 \sum_{i}(|z_i|^2 + |w_i|^2)),
\label{gHR}
\end{eqnarray}
This wave function can be generated by an electron operator whose non-abelian
part is $\partial^l \psi_{e,\sigma}$. 
We would like to point out that only for $l=0$
can ${\cal A}_{z,w}(\frac{1}{(z_1-w_1)^{2l+2}}
\frac{1}{(z_2-w_2)^{2l+2}}...)$ be a correlation of a primary field of a CFT.
For other $l$, it is a correlation of descendant field.
The operator that produces a cut of $-1$
to $\psi_{e,\sigma}$ also produces a cut of $-1$ to $\partial^l \psi_{e,\sigma}$.
Thus, for generalized HR state, although the electron exponent
is changed to $g_e=m+2l+2$, the quasiparticle exponent remains the same
$g_q=\frac{1}{4}+\frac{1}{4m}$. If we calculate the quasiparticle correlation
using the first-quantized picture, we find the non-abelian 
part has a correlation $<0|U(x_1,x_2)|0>=\hbox{det}
\big(\left(\frac{m}{n}\right)^{l+\frac12} f_{nm}\big)$
which is independent of $l$ as expected.
This provides an important consistency check of the formalism we introduce 
in this paper. We will see next that the similar behavior is also shared by
several other non-abelian QH states.

Now let us consider two other types of non-abelian states.
The first one is the generalized Pfaffian wave function\cite{a}
\begin{eqnarray}
\Phi_p &=& \Phi_{Pf} \Phi_m , \nonumber\\
\Phi_{Pf} &=& {\cal A}{\frac{1}{(z_1-z_2)^{2l+1}}\frac{1}{(z_3-z_4)^{2l+1}}...},
\nonumber\\
\Phi_m &=& ( \prod_{i<j} (z_i-z_j)^m ) \exp(-\frac14 \sum_{i}|z_i|^2),
\label{25}
\end{eqnarray}
where ${\cal A}$
is the anti-symmetrization operator; $l$ is an integer and $ m $ is an
even integer. The second one is the generalized d-wave paired FQH states for
spinless electrons\cite{g}
\begin{eqnarray}
\Phi &=& \Phi_d \Phi_m ,
\nonumber\\
\Phi_d &=& {\cal S}{\frac{1}{(z_1-z_2)^{2l+2}} \frac{1}{(z_3-z_4)^{2l+2}}...},
\nonumber\\
\Phi_m &=& ( \prod_{i<j} (z_i-z_j)^m ) \exp(-\frac14 \sum_{i}|z_i|^2),
\label{26}
\end{eqnarray}
where ${\cal S}$
is the symmetrization operator; $ l $ is an integer and $ m $ is an odd
integer.
In the CFT approach, the non-abelian parts of electron operators for states
$(\ref{25})$ and $(\ref{26})$ are identified as $\partial^l \psi$, where $\psi$
is a dimension 1/2 primary field in a $c = 1/2$ 
CFT for state $(\ref{25})$ and is
a dimension 1 primary field in a $c = 1$ CFT for state 
$(\ref{26})$. For the case of $l = 0$ $\cite{f}$, the quasiparticle operator 
$ \eta$  satisfies the following operator product expansion (OPE)
\begin{equation}
\psi(z) \eta(0) \sim z^{-1/2} [ \mu(0) + O(z)],
\label{27}
\end{equation}
where $\mu$ is another operator. For the case of general $ l$, we take $
l$-times derivatives on Eq. $(\ref{27})$. It is easy to see that the
cut and the
dimension of the quasiparticle operator $\eta$ and hence the exponent $g_q$ will
not be affected by the differentiation. Both states have charge
$1/2m$ quasiparticles. The electron and quasiparticle exponents are 
$(g_e,g_q)=(m+2l+1,\frac18+\frac{1}{4m})$ 
for state (\ref{25}), and $(g_e,g_q)=(m+2l+2,\frac18+\frac{1}{4m})$
for state (\ref{26}).\cite{g}

The third example is the generalized d-wave-paired-spin-triplet 
FQH state for spin-1/2 electrons\cite{g}
\begin{eqnarray}
\Phi_{DL} (z_i,w_i) &=& \Phi_{kmn}\Phi_{dt}(z_i,w_i),
\nonumber\\
\Phi_{dt} &=&{\cal S}_{z,w}{ \frac{1}{(z_1-w_1)^{2l+2}}
\frac{1}{(z_2-w_2)^{2l+2}}...},
\nonumber\\
\Phi_{m} &=& { \prod_{i<j} (z_i-z_j)^m \prod_{i<j} (w_i-w_j)^m \prod_{i<j}
(z_i-w_j)^m } \exp(-\frac14 \sum_{i}(|z_i|^2 + |w_i|^2)),
\label{28}
\end{eqnarray}
where ${\cal S}_{z,w}$ is the symmetrization operator which performs separate
symmetrizations among $z_i$'s and among $w_i$'s; $l$ is an integer and $m$ is an
odd integer. For the case of $l = 0$ $\cite{g,f}$, the quasiparticle operator
$ \eta$  satisfies the following operator product expansion (OPE)
\begin{equation}
\psi_{\pm}(z) \eta(0) \sim z^{-1/2} [ \mu_{\pm}(0) + O(z)],
\label{28-1}
\end{equation}
where the non-abelian part of the electron operators $ \psi_{\pm}$ were
identified as the currents of a $ U(1)\times U(1)$ Gaussian model which is
a $c = 2$
CFT, and  $\mu_{\pm}$ are some other operators.  For the case of general $l$,
we take $ l$-times derivatives on Eq. $(\ref{28-1})$ as we did before.
The charge $1/2m$ quasiparticle has an exponent
$g_q=\frac14 +\frac{1}{4m}$.\cite{g} The electron exponent is $g_e=m+2l+2$.

\section{Summary}

In this paper we calculated the exponent $g_q=\frac14+\frac{1}{4m}$ 
of the quasiparticle propagator $1/(x-vt)^{g_q}$ for the HR state with filling
factor $\nu=1/m$. Our calculation is based on two assumptions:
\begin{itemize}
\item
The edge spectrum (the non-abelian part) is described by character
Eq.~(\ref{1}). This is checked and confirmed (to certain low levels)
by numerical calculations.
\item
The non-abelian part of electron operator has a propagator
$<0|\psi_{e,\sigma}\psi_{e,\sigma'}|0>=\delta_{\sigma \sigma'}/
(x-vt)^2$. This assumption is checked by the numerical results
presented in this paper.
\end{itemize}
The two assumptions allow us to construct the Hamiltonian and commutators
in Eq.~(\ref{2}).

\acknowledgments
JCL is supported by NSC grant of Taiwan, R.O.C.
XGW is supported by NSF grant DMR-94-11574
and A.P. Sloan fellowship.

\end{document}